\documentclass[useAMS,usenatbib]{mn2e}

\usepackage{graphicx}

\title[Photonic spatial reformatting of stellar light for diffraction-limited spectroscopy]
  {Photonic spatial reformatting of stellar light for diffraction-limited spectroscopy}
\author[R. Harris et al.]
  {R. J. Harris,$^1$\thanks{These authors contributed equally to this work}\thanks{Email: r.j.harris@durham.ac.uk}
   D. G. MacLachlan,$^2$\footnotemark[1]\thanks{Email: dgm4@hw.ac.uk}
   D. Choudhury,$^2$
   T. J. Morris,$^1$
   E. Gendron,$^3$ \newauthor
   A. G. Basden,$^1$
  G. Brown,$^4$
  J. R. Allington-Smith$^1$ and
  R. R. Thomson$^2$\\
  $^1$Department of Physics, University of Durham, South Road, Durham, DH1 3LE, UK\\
  $^2$SUPA, Institute of Photonics and Quantum Sciences, Heriot-Watt University, Edinburgh, EH14 4AS, UK\\
   $^3$LESIA, Observatoire de Paris, Meudon, 5 Place Jules Janssen, 92195 Meudon, France\\
   $^4$Optoscribe Ltd., Kirkton Campus, Livingston, EH54 7DQ, UK}
\date{2014 Xxxxx XX}

\pagerange{\pageref{firstpage}--\pageref{lastpage}} \pubyear{2014}

\def\LaTeX{L\kern-.36em\raise.3ex\hbox{a}\kern-.15em
    T\kern-.1667em\lower.7ex\hbox{E}\kern-.125emX}

\begin{document}

\label{firstpage}

\maketitle

\begin{abstract}
 The spectral resolution of a dispersive spectrograph is dependent on the width of the entrance slit. This means that astronomical spectrographs trade-off throughput with spectral resolving power. Recently, optical guided-wave transitions known as photonic lanterns have been proposed to circumvent this trade-off, by enabling the efficient reformatting of multimode light into a pseudo-slit which is highly multimode in one axis, but diffraction-limited in the other. Here, we demonstrate the successful reformatting of a telescope point spread function into such a slit using a three-dimensional integrated optical waveguide device, which we name the photonic dicer. Using the CANARY adaptive optics demonstrator on the William Herschel Telescope, and light centred at 1530~nm with a 160~nm FWHM, the device shows a transmission of between 10 and 20\% depending upon the type of AO correction applied. Most of the loss is due to the overfilling of the input aperture in poor and moderate seeing. Taking this into account, the photonic device itself has a transmission of 57 $\pm$ 4\%.We show how a fully-optimised device can be used with AO to provide efficient spectroscopy at high spectral resolution.
\end{abstract}


\section{Introduction}\label{intro}

Spectroscopy is a technique of paramount importance to modern astronomy. Analysis of spectroscopic data can reveal key information including the chemical composition, distance, velocity, temperature and density of astronomical sources (e.g. \citealt{Kwok2004}, \citealt{Swain2010},  \citealt{Eisenhauer2003}, \citealt{Ghez2003}, \citealt{Chauvin2004},  \citealt{Marois2008} and references therein). The most commonly used spectrograph configuration in astronomy is the dispersive spectrograph. In its simplest form, such a spectrograph consists of a slit, into which light from the object is coupled, an optic such as a lens or mirror to collimate the light from the slit, a dispersive element such as a grating or prism to induce chromatic angular dispersion, and a further optic to image the slit to the detection plane. In such a spectrograph, the position of the slit image at the detection plane becomes a function of wavelength, which can be recorded using a detector array. The resolving power of such a spectrograph ($R= \lambda / \delta\lambda$, where $\lambda$ is the measurement wavelength and $\delta\lambda$ the minimum resolvable wavelength difference) varies inversely with the slit width, and the maximum resolving power of a given instrument is obtained when the slit width is matched to the diffraction-limit of the collimator. As a consequence, a larger slit width is required to efficiently accept multimode light into the spectrograph and this will result in a reduced spectral resolving power for a given diameter of collimated beam. In order to increase the spectral resolving power a larger collimated beam and hence grating are required, increasing the size and cost of the instrument.

The trade-off between spectral resolving power and the coupling efficiency of multimode light into a dispersive spectrograph is a key design aspect of modern astronomical spectrographs. As the number of spatial modes that form the telescope point-spread-function (PSF) increases with the square of the telescope aperture ($D_T$) \citep{Harris2012}, the spatial element size and hence spectrograph size for a given resolving power also increases. For the current generation of 8-10~m class telescopes, this scaling is already resulting in spectrographs with volumes of many cubic meters, a trend that will only continue as Extremely Large Telescopes (ELTs) with apertures greater than 30~m diameter are constructed \citep{Cunningham2009}.

One approach to address this increase in spectrograph size is to use Adaptive-Optics (AO). A perfect AO system would provide a diffraction-limited image regardless of telescope size, meaning that, in principle, the spectrograph size would also be independent of the telescope diameter. However, the complexity of the AO system required to achieve a given level of performance also scales with the square of $D_{T}$. Furthermore, AO performance is also highly dependent on the type of correction being applied and the strength of atmospheric turbulence. This is particularly true if the AO system is providing correction at visible wavelengths, where currently proposed ELT AO systems can only provide a partially-corrected PSF.

A second approach to reduce spectrograph size is to reformat the image into multiple slices aligned in a slit, a technique known as  image slicing e.g. \citet{Weitzel1996} and references therein. These slices can then be distributed between numerous replicated instruments as described in \citet{Hill2008}. Whilst this approach still results in a similar overall instrument volume the individual instruments can be made smaller \citep{Harris2012}.

\citet{Bland-Hawthorn2010} proposed that optical fibre guided-wave transitions could be used to efficiently reformat the multimode telescope PSF into a diffraction-limited (in one axis) pseudo-slit, which can then be used as the input to a dispersive spectrograph. This approach has the potential to enable the efficient use of diffraction-limited spectrographs on large optical or near-infrared telescopes \citep{Cvetojevic2009,Cvetojevic2012}. The use of single-mode (or diffraction-limited) waveguides also has the potential to reduce or eliminate modal noise (again in one axis). This is a major limiting factor in the performance of current high resolution fibre-fed spectrographs \citep{Lemke2011}.  

The key to this vision is the photonic lantern, a remarkable new class of optical guided-wave transition that facilitates the low-loss coupling of multimode light into an array of single-modes. After the photonic lantern transition, these single-modes can be re-arranged to produce a diffraction-limited slit, or pseudo-slit, consisting of a linear array of single-modes.

Photonic lanterns can currently be fabricated using two different methods. The first relies on the application of optical fibre tapering techniques, using either bundles of single-mode fibres (\citealt{Leon-Saval2005}, \citealt{Noordegraaf2009}), or a single multicore fibre (MCF) \citep{Birks2012}. Once the discrete single-modes are generated by the photonic lantern, they can be re-arranged into a diffraction-limited pseudo-slit. This can be achieved either by re-arranging the individual single-mode fibres themselves in the case of fibre-bundle type photonic lanterns (\citealt{Bland-Hawthorn2010}, \citealt{Bland-Hawthorn2011}), or by using a three-dimensional reformatting component in the case of MCF-type photonic lanterns \citep{Thomson2012,Spaleniak2013}. If the MCF itself only supports a low number of modes, it is also possible to use each core as an individual diffraction-limited slit, and very recently \citet{Betters2013} have demonstrated the acquisition of a solar spectrum using a diffraction-limited spectrograph based on this concept. The second fabrication route relies on the application of ultrafast laser inscription (ULI), a laser based fabrication technology where focused ultrashort laser pulses are used to directly write three-dimensional optical waveguide structures into a transparent substrate material e.g. \citet{Davis1996}, \citet{Nolte2003}, \citet{Gatass2008}, \citet{Thomson2009}, \citet{Thomson2011}, \citet{Jovanovic2012}. Previously, we have proposed that ULI could be used to fabricate an integrated optical waveguide device that seamlessly combines photonic lantern and reformatting functions onto a single chip \citep{Thomson2009} â a device we will call a ''photonic dicer" since it "dices" the PSF in a manner similar to the way conventional bulk-optic image slicers operate, but unlike image slicers, does not preserve spatial information. This fully integrated dicer solution uniquely enables the N single-mode waveguides to be reformatted into a slab waveguide, which is single-mode in one axis, but multimode in the other. In this case, all N spatial modes along the pseudo-slit are spatially overlapped, and any modal noise in the input is converted into amplitude and phase variations along the slit. This minimises the length of the slit, and will minimise the number of detector pixels (and hence detector noise) required to analyse the light in the spectrograph.

In this paper, we report the successful application of an integrated photonic dicer to the reformatting of an AO-corrected telescope PSF into a diffraction-limited pseudo-slit. The integrated reformatting device presented here offers an alternative method to achieve high-resolution spectroscopy of astronomical sources. In Section~\ref{dicer} we describe the design and fabrication of the photonic dicer, followed by a description of the experimental setup at the telescope in Section~\ref{canary}. Then we present results and analysis of the laboratory and on-sky testing of the photonic dicer in Section~\ref{results}. Finally we present a short summary of the results and point to future work in Section~\ref{conc}.

\section{Photonic dicer design}\label{dicer}

\begin{figure*}
 \includegraphics[width=170mm]{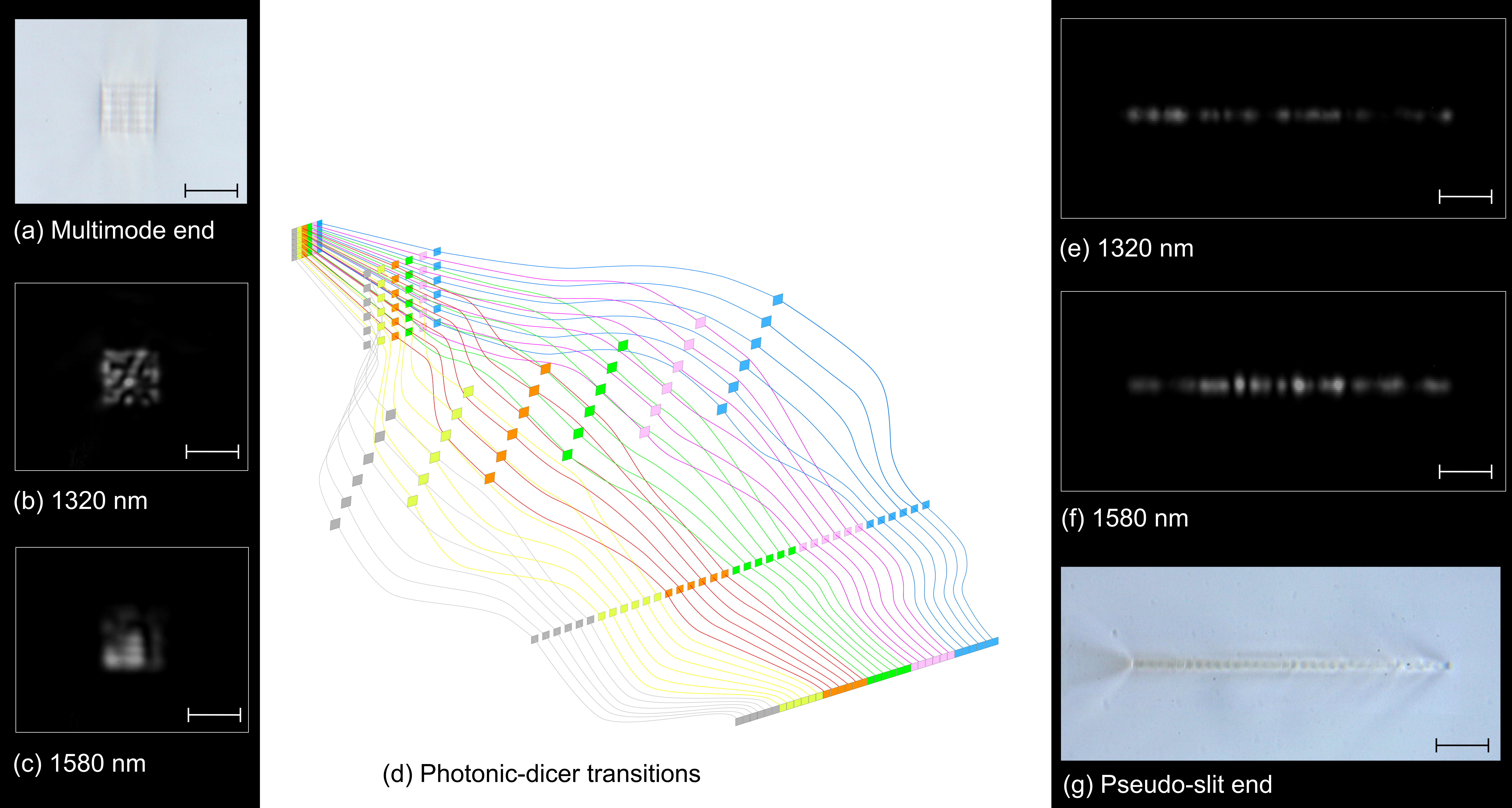}
 \caption{The design of the integrated photonic dicer. (a) Transmission microscope image of the $6\times6$ array multimode input facet where the telescope PSF is injected. Mode image of input facet when excited using monochromatic light at (b) 1320~nm and (c) 1580~nm. (d) Schematic diagram showing the colour coded trajectories of the optical waveguides that constitute the photonic dicer. Mode image of the pseudo-slit when the input facet is injected using monochromatic light at (e) 1320~nm and (f) 1580~nm. (g) Transmission microscope image of the pseudo-slit where the reformatted diffraction-limited output is formed. Scale bar for each image represents 50~$\umu$m.}
 \label{fig1}
\end{figure*}

The ideal photonic dicer would collect the multimode PSF from the telescope, transform it into a two-dimensional array of single-modes using a photonic lantern transition, reformat this two-dimensional array of single-modes into a one-dimensional array and finally bring these together to form a diffraction-limited pseudo-slit, all with zero loss. In reality, however, the losses incurred in a device due to the transitions and the spatial reformatting are dependent on many factors, including bend, waveguide, propagation and radiation losses due to non-adiabatic transitions. The coupling loss between the telescope and the photonic dicer on the other hand is determined by the degree of matching between the shape and number of modes supported by the multimode entrance to the photonic dicer and those that form the PSF.
The number of modes that compose a telescope PSF can be approximated \citep{Harris2012} by  

\begin{equation}
  M\approx(\pi\chi D_T/4\lambda)^2			
  \label{Eq1}
\end{equation}

where $D_T$ is 4.2~m for the WHT, $\chi$ is the angular size of the PSF in radians, usually defined by the full-width at half-maximum (FWHM) intensity of the PSF. The central wavelength ($\lambda$) of our on-sky measurements was 1530~nm \citep{Vidal2014}. This was determined from the overlap between the passband of the H-band filter and the responsivity of the Xenics Xeva-1.7 320 InGaAs camera used as a NIR imager for the experiment. From Eq.~\ref{Eq1}, it is clear that the modal content of the telescope PSF scales directly with changes in astronomical seeing or the telescope diameter \citep{Harris2012}. AO reduces this dependence by correcting for the turbulence in the atmosphere, effectively reducing the size of the PSF. This consequently reduces the number of modes contained in the PSF.

CANARY is an on-sky AO demonstrator system \citep{Myers2008} installed on the 4.2~m William Herschel Telescope (WHT) in La Palma, Spain. CANARY is a low-order system that was developed to principally investigate novel AO instrumentation and control techniques and cannot provide truly diffraction-limited imaging except under exceptional atmospheric conditions. Under median seeing conditions, with CANARY operating in closed-loop mode, a typical AO corrected H-band (approximately 1490~nm to 1790~nm) PSF has a Strehl Ratio of 20-30\% \citep{Gendron2011}. Using previously recorded AO-corrected PSFs from CANARY we calculated that the PSF could in principle, be efficiently coupled into the multimode end of a photonic dicer supporting 5 spatial modes along each axis.

For the on-sky work reported here, however, we deliberately chose to use a photonic dicer with a multimode entrance supporting approximately 6 spatial modes in each axis. This decision was made to increase alignment tolerances and minimise any potential losses due to spatial mode number and shape mismatch between the photonic dicer and the PSF. The multimode end of the photonic dicer was thus designed to be approximately 50~$\umu$m$\times$50$~\umu$m, while the pseudo-slit was designed to be approximately 300~$\umu$m long and 9~$\umu$m wide. The photonic dicer design is shown in Fig.~\ref{fig1}, where it can be seen that the waveguides undergo a number of distinct transition steps. First, the multimode waveguide splits, in a symmetric fashion, into a 6$\times$6 array of uncoupled waveguides that were measured to be single-mode throughout the 1320~nm to 1580~nm spectral region. This is the photonic lantern transition. The now separate single-mode waveguides then undergo three further reformatting stages to finally bring them together to form a slab waveguide (the pseudo-slit), which is single-mode in one axis, but highly multimode in the other.

The substrate used to fabricate the photonic dicer structures was a borosilicate glass (Corning, EAGLE 2000). Optical waveguides were inscribed using 460~fs pulses from a Fianium Femtopower 1060 femtosecond laser, operating at repetition rate of 500~kHz and a central wavelength of 1064~nm. The waveguides were inscribed in the transverse inscription geometry using the multiscan technique \citep{Nasu2005}. Each single-mode waveguide was inscribed using 21 scans, with each scan separated by an offset of 0.4~$\umu$m. The pulse energy used was 251~nJ, with the pulses focused at a depth of approximately 200~$\umu$m beneath the surface using a 0.3 NA aspheric lens. The inscription beam was circularly polarised. The substrate was translated through the beam focus at a speed of 8~mms$^{-1}$ using cross roller-bearing xyz translation stages (Aerotech ANT130). Fig.~\ref{fig1}(a) shows the multimode entrance to our photonic dicer that is constructed by bringing 36 single-mode waveguides together with a centre-to-centre separation of 8.4~$\umu$m to form a square multimode waveguide supporting approximately 6 spatial modes in each orthogonal axis. Fig.~\ref{fig1} also shows the pseudo-slit output of the photonic dicer. This is constructed by reformatting the 36 waveguides into a single linear array with a centre-to-centre separation of 8.4~$\umu$m. Full fabrication details and characterisation of the photonic dicer can be found  in \citealt{MacLachlan2014}.

\section{CANARY setup and integration}\label{canary}
For the work presented here, CANARY was configured to provide closed-loop AO correction using an on-axis natural guide star as a wavefront reference. Light with a wavelength $>$1000~nm was sent to the multimode entrance of the photonic dicer with visible wavelengths directed to a 36 sub-aperture Shack-Hartmann Wavefront Sensor (WFS). The WFS measurements were used to drive a 56-actuator deformable mirror and a separate fast-steering mirror which could provide a partially corrected PSF at a wavelength of 1500~nm. The WFS was placed behind the deformable mirror measuring the residual wavefront error after correction. The deformable mirror commands were calculated using a basic integrator feedback controller with a closed-loop gain of 0.3.

\begin{figure*}
 \includegraphics[width=170mm]{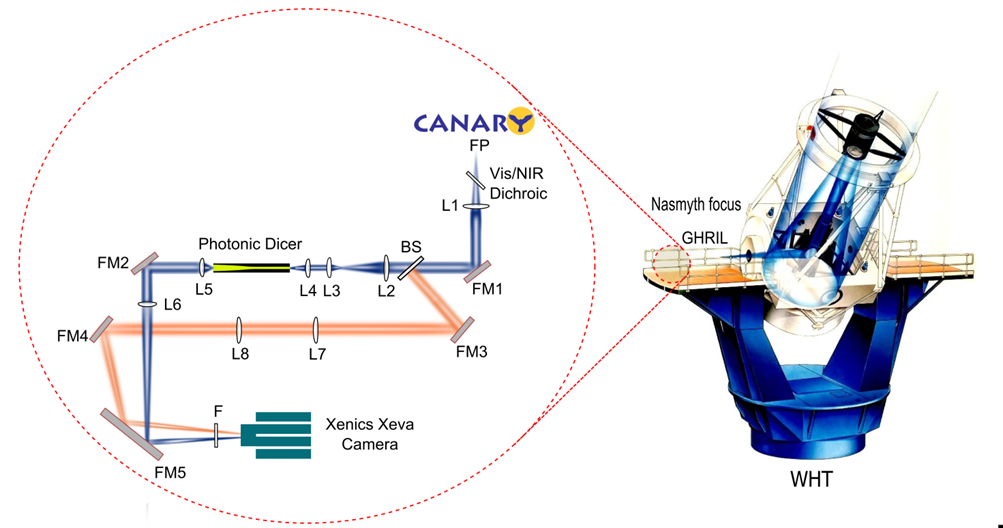}
 \caption{Schematic diagram showing the free-space optical setup used for the on-sky tests. Stellar light collected by the William Herschel Telescope is fed through the CANARY Adaptive Optics system which generates a corrected multimode PSF. A dichroic is used to remove the visible part of the spectrum ($<1000$~nm) from the light beam, which is then collimated using a lens (L1). Using a beamsplitter (BS), 10\% of the collimated beam is fed to a reference path, which re-images the PSF onto the camera using lenses L7 and L8. The remaining light is focused using a lens (L2) and subsequently passes through a telescope (lenses L3 and L4), which allows the PSF to be injected into the input facet of the photonic dicer. The reformatted output from the device is imaged onto the camera using lenses L5 and L6. The fold mirrors (FM1 - FM5) are utilised for beam steering. (WHT image courtesy of the Isaac Newton Group of Telescopes, La Palma).}
 \label{fig2}
\end{figure*}

CANARY includes a range of sources that can be used when not on-sky to calibrate and optimise system performance. For our purposes, a 1550~nm laser was coupled into a single-mode fibre and placed at the input focal plane of CANARY. Light from this calibration source passes through the CANARY optical train (including the deformable mirror and fast-steering mirror) and the output PSF is re-imaged onto the photonic dicer multimode entrance using the setup shown in Fig.~\ref{fig2}. The PSF at the multimode input of the photonic dicer was then modified by changing the surface shape of the deformable mirror to maximise the detected flux from the pseudo-slit output. To modify the PSF, the AO-correction loop was engaged and static offset terms were artificially applied to the measured WFS signals. These offset terms were automatically adjusted through a process of simulated annealing \citep{Kirkpatrick1983}. The final WFS shape that provided maximum signal at the pseudo-slit output was recorded and used as a reference for all subsequent on-sky operations.

CANARY was operated in three modes during data acquisition to investigate the photonic dicer performance under different degrees of AO correction. In closed-loop mode, CANARY provided correction of both tip-tilt and higher-order wavefront aberrations at an update rate of 150~Hz. Tip-tilt mode reduced the integrator loop gain on the high-order modes to a small value (typically 0.001). In this way only PSF location was stabilised in real-time but the optimised PSF shape (determined via the simulated annealing method during calibration) was maintained without providing high-order AO correction. Open-loop mode was implemented by additionally reduced the gain on the tip-tilt correction to maintain the PSF in the reference location for optimum coupling without providing high temporal frequency correction.

The setup shown in Fig.~\ref{fig2} was designed to produce an image of the PSF by coupling approximately 10\% of the light directly to the InGaAs camera without passing through the photonic dicer. This reference path enables us to monitor the PSF shape and input flux directly, and therefore evaluate the transmission of the photonic dicer regardless of the PSF variation. The remaining fraction of the light is coupled into the photonic dicer, and the pseudo-slit output is imaged to the camera. The combination of the InGaAs camera response and H-band filter placed in front of the camera defined a detection bandwidth of 160~nm (FWHM) \citep{Vidal2014}. Linearity of the InGaAs detector was tested in the laboratory with a 1550~nm LED and an integrating sphere with a calibrated power meter. The response was found to be linear to within $<<$1\%. 

The telescope provides an F/11 beam with a plate scale value of $4.54\arcsec$mm$^{-1}$, but this is converted by the input coupling optics (L3 and L4 in Fig.~\ref{fig2}). The experiment was designed produce an F/7.3 beam with a plate scale of $6.88\arcsec$mm$^{-1}$. This means that the 50~$\umu$m$\times$50$~\umu$m multimode end of the photonic dicer corresponds to a spatial scale of $0.35\arcsec$ on-sky in each axis. However optimal coupling was achieved during tests with a beam of F/6.2 giving a plate scale of $7.96\arcsec$mm$^{-1}$ meaning the angular size of the photonic dicer was $0.405\arcsec$.

Each dataset captured consisted of $\approx200$ NIR images, each of which was obtained by exposing the camera for 400~ms, in order to acquire an adequate signal without saturation. Once the datasets were acquired with the photonic dicer in place, the device was then removed and the PSF from the input coupling plane was imaged directly onto the camera by translating L5 along the collimated beam. A set of $\approx100$ images was then acquired in order to calibrate the difference in integrated power between each path. Use of light from the star itself in this manner allows calibration that accounts for device transmission (the ratio of input flux to output flux) of the system and stellar type. A series of dark and sky background images were also acquired in order to determine the background noise floor.

\section{Results}\label{results}
The photonic dicer was tested on-sky at the WHT on the 13th of September 2013, and all data was acquired between 21:45 and 22:30 GMT. The star selected for observation was TYC1052-3027-1 from the Tycho 2 catalogue \citep{Hog2000}, a $3^{rd}$ magnitude star in the astronomical H-band. The astronomical seeing values, as measured using an on-site monitor \citep{OMahoney2003}, varied between $0.7\arcsec$ and $0.95\arcsec$ over the course of the measurements, these are representative of median to poor seeing for the telescope site.

\begin{figure}
 \includegraphics[width=84mm]{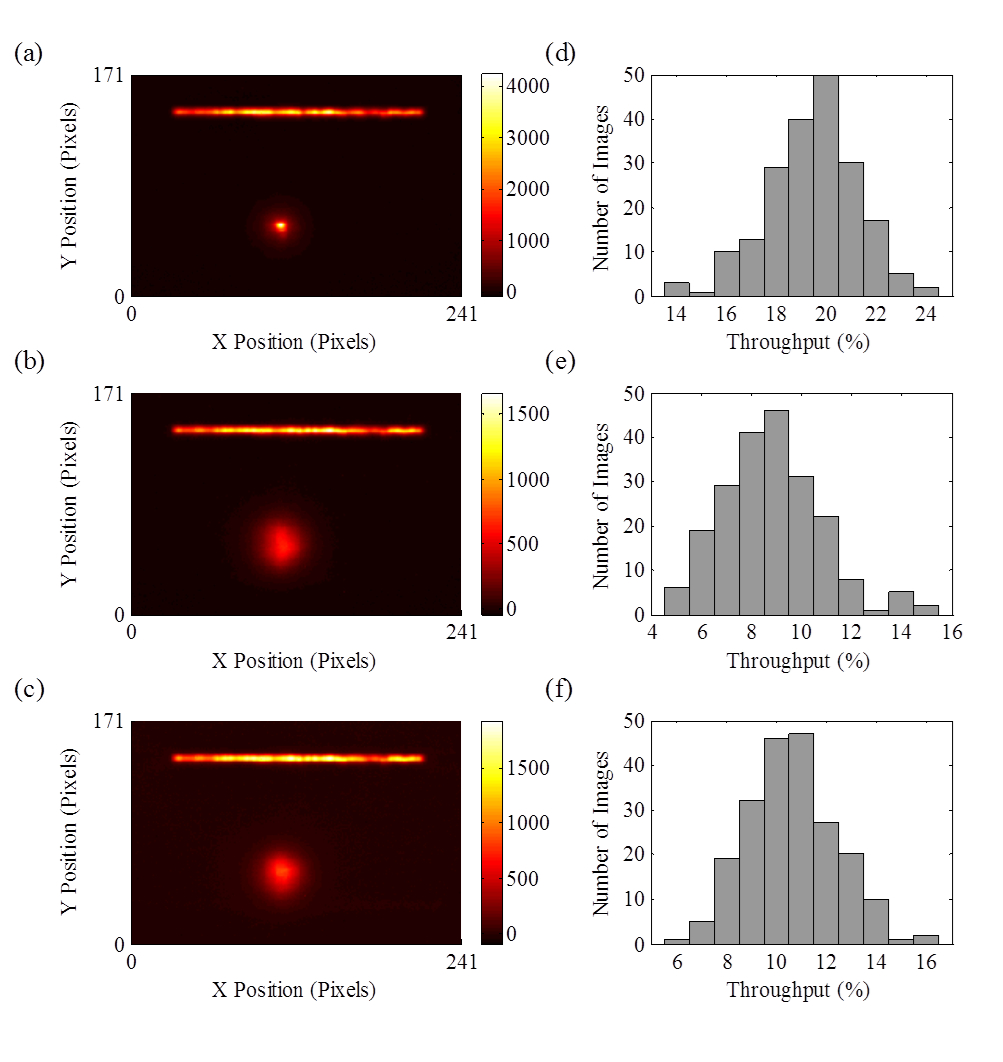}
 \caption{(a-c) Colour map images of the camera sensor showing the reformatted output of the photonic dicer and the telescope PSF from the reference path for (a) closed-loop, (b) tip-tilt only and (c) open-loop modes of AO correction. The images show the average reduced data for 200 frames for each AO mode. Note: The camera has $30~\umu$m pixels and the images for each AO mode have different intensity scales. (d-f) Histogram plots showing percentage transmission distribution over the number of images acquired for (d) closed-loop, (e) tip-tilt and (f) open-loop AO modes. Hot pixel removal and background correction algorithms have been applied prior to evaluating the transmission for each image frame acquired.}
 \label{fig3}
\end{figure}

Fig.~\ref{fig3} shows the averaged images of the pseudo-slit and PSF captured when the AO system was operating in closed-loop, tip-tilt only and open-loop modes respectively. It should be highlighted that each of the images has a different intensity scale, with the detected flux under closed-loop operation being a factor $\approx$~2 higher than that measured under the other two AO modes.

In the case of tip-tilt correction, the photonic dicer transmission was measured to be 9\%$\pm$2\%. In open-loop operation, the transmission was measured to be 10.5\%$\pm$2\%, the higher value being due to better seeing conditions at the time the observations were taken. In the case of closed-loop operation, with full AO correction, the transmission of the photonic dicer was measured to be 19.5\%$\pm$2\%. Histograms of the transmission data obtained for closed-loop, tip-tilt and open-loop AO modes are also shown in Fig.~\ref{fig3}. 

\begin{figure}
\includegraphics[width=84mm]{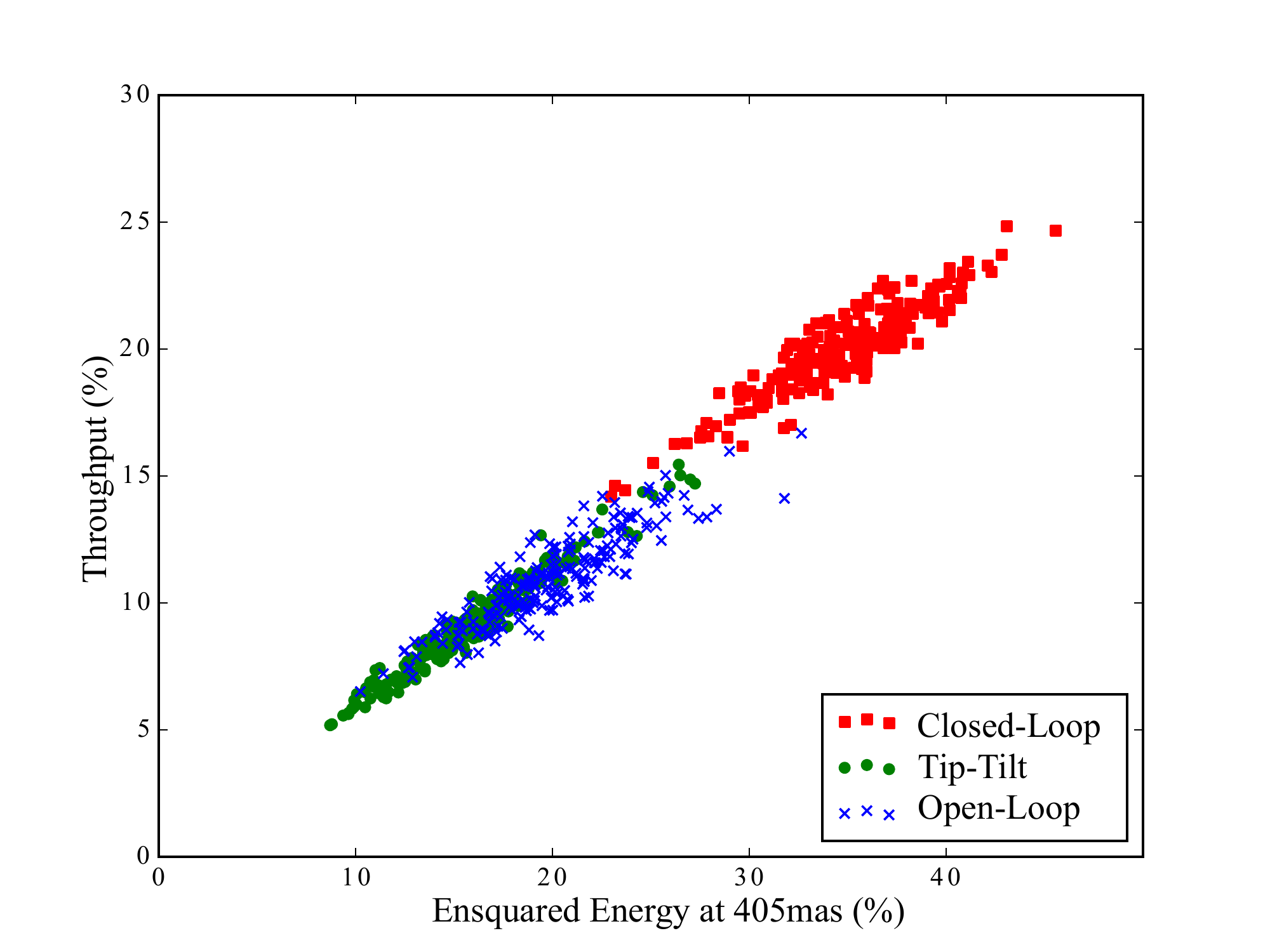}
 \caption{The measured transmission from the slit end of the photonic dicer plotted against the ensquared energy from the reference arm at 405mas. This is shown for the three AO operating modes.}
 \label{fig4}
\end{figure}

\begin{figure}
\includegraphics[width=84mm]{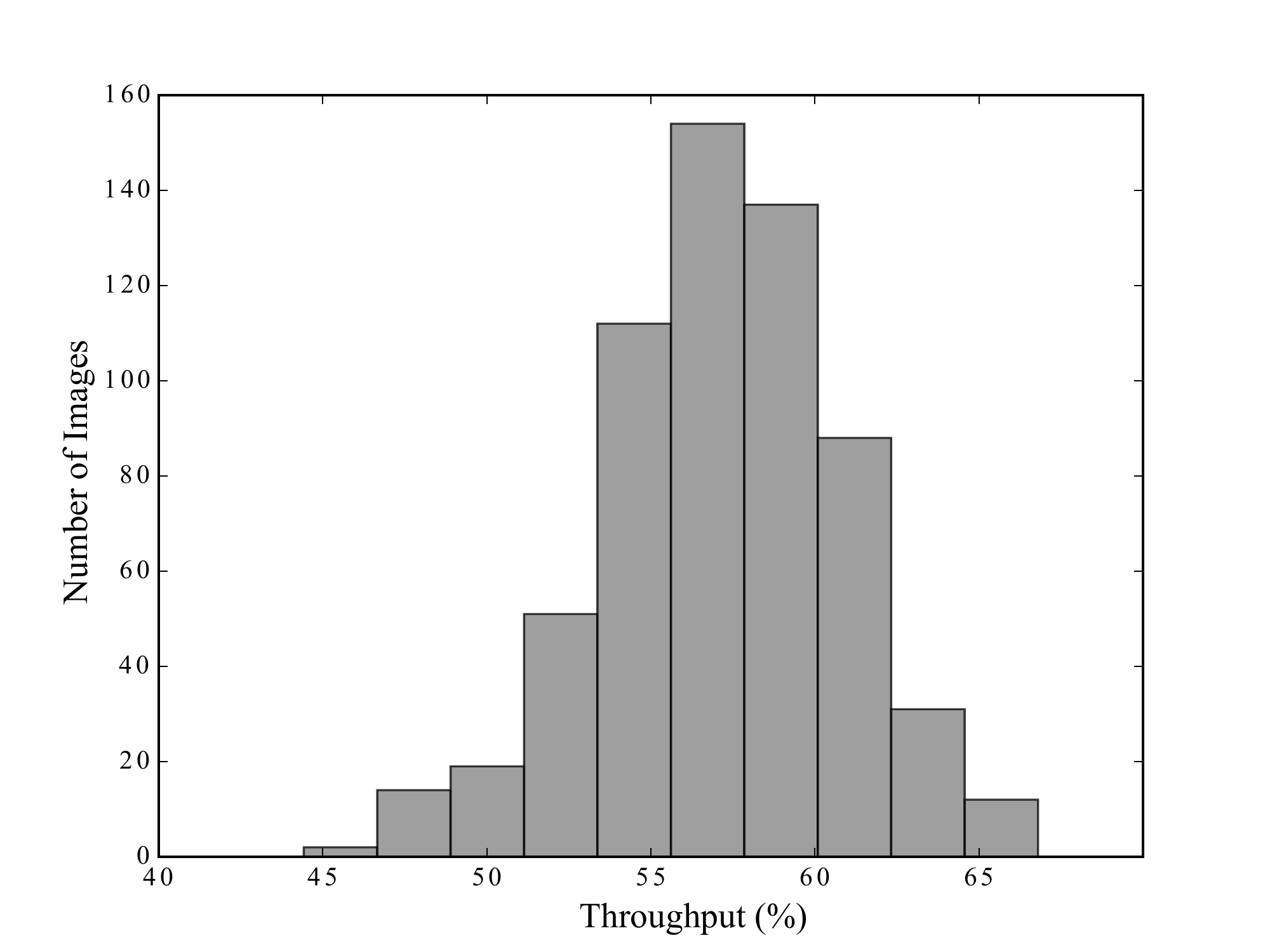}
 \caption{A histogram of the estimated transmission of the photonic dicer for all AO modes.  This is calculated by dividing the calculated transmission from the slit end of the photonic dicer by the ensquared energy at 405mas in the reference arm.}
 \label{fig5}
\end{figure}

In order to quantify where the losses occurred the images were averaged and the centre of the dicer located from the averaged maximum of the reference image. By taking the ensquared energy using this location the energy coupled to the photonic dicer in each image was calculated. This was compared to the calculated transmission and the results are shown in Fig.~\ref{fig4}. They show a linear increase in transmission with increase in ensquared energy. By dividing the measured transmission by the calculated ensquared energy the on-sky transmission of the photonic dicer can be estimated as $57\pm 4$\%. A histogram of this is shown in Fig.~\ref{fig5}. The photonic dicer transmission was also measured in the laboratory, using monochromatic multimode laser light, in a manner similar to that described by \citet{Thomson2011}. These laboratory measurements indicated that a maximum transmission of 66\% $\pm$ 3\% could be achieved for light in the 1320~nm to 1580~nm spectral region. 

To utilise the photonic dicer as an input for a high-resolution astronomical spectrograph, the pseudo-slit should ideally be as uniform as possible, in terms of its width and the centering of the modes along the slit (perpendicular to the slit length). This stability is most important with time, as any changes cannot be calibrated out.

\begin{table}
 \centering
 \begin{minipage}{84mm}
  \caption{The measured instantaneous centring and slit width variations before average subtraction. The first set of results was performed in laboratory conditions using a stable source. The on-sky results have a bandwidth 160~nm (FWHM) centred at 1530~nm with a varying input.}
	\begin{tabular}{ | c | c | c |}  \hline 
		Wavelength  & Slit Centring & Slit Width   \\
		 (nm) & (\% of FWHM) & (\% of FWHM) \\ \hline
		1320 & 14 & 16  \\
		1400 & 16 & 14  \\
		1500 & 12 &  18 \\
		1550 & 14 & 20  \\
		1580 & 13 &  18 \\ \hline 
		On-sky & 15 & 20 \\ \hline
	\end{tabular} 
	\label{tab:slit_params}
\end{minipage}
\end{table}

Summing in time and over position along the slit, the central location of the slit (in the dispersion direction) was found to vary by a maximum of $\pm$0.35 pixels. The mean of the fitted Gaussian width was 4.55 pixels with a maximum variation of $\pm$0.5 pixels. The variation with time for the same position along the slit was determined  from a histogram of all instantaneous locations with the time-averaged value for that position along the slit subtracted.  
 The FWHM of the distribution of the variation of slit locations with time in the dispersion direction was found to be 0.05 pixels. The variation of slit-width was determined in a similar fashion from a histogram of instantaneous slit widths with the time-averaged width subtracted for that location along the slit. The FWHM of the distribution of relative slit widths was found to be 0.04 pixels.

This measurement was also repeated in the laboratory with monochromatic laser light and the positional variations as percent of the average Gaussian FWHM are shown in Tab.\ref{tab:slit_params}. These measurements show the photonic dicer is stable with time in the dispersion axis, a vital feature for a high resolution spectrograph. They also show the need for high tolerances in manufacture in any future devices, as any variation in slit parameters will result in a potential reduction of spectral resolution.

\begin{figure}
\includegraphics[width=84mm]{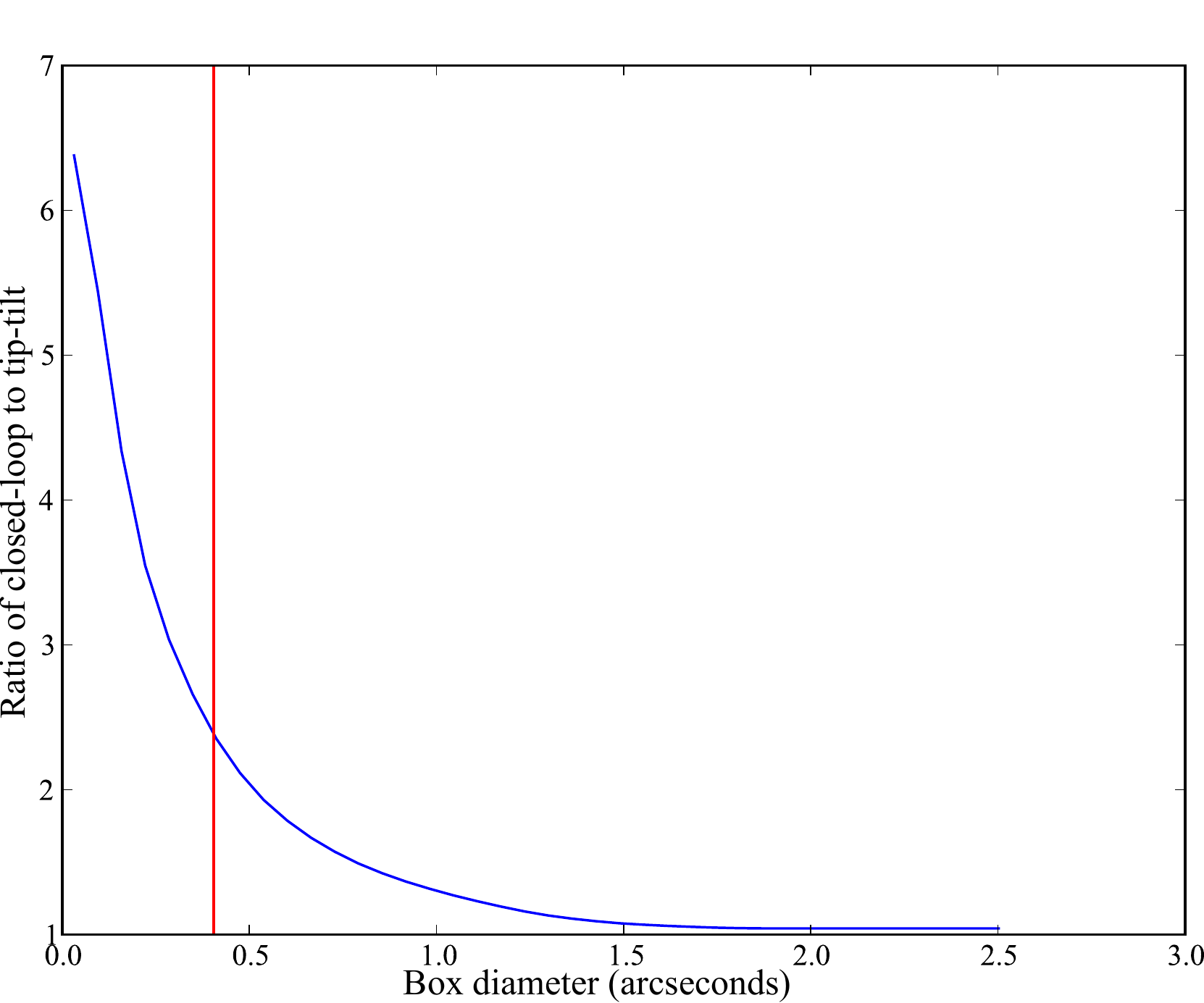}
 \caption{The ratio of ensquared energy between closed-loop and tip-tilt only AO modes as a function of photonic dicer input size based on analysis of the AO-corrected PSFs from CANARY. The vertical red line indicates the spatial sampling size of the photonic dicer used on-sky.}
 \label{fig6}
\end{figure}

In order to optimise the performance of the system, we attempted to match the FWHM of the PSF from CANARY to our photonic dicer to achieve efficient coupling whilst restricting the size of the device. Future instruments will be required to balance the number of modes (and hence slit length) against the desire for maximal transmission of the system. Key to this will be the level of AO correction required to attain a given transmission for a given number of modes. Fig.~\ref{fig6} shows the relation between the ensquared energy under closed-loop and tip-tilt correction for our averaged images on CANARY. For our chip, under closed-loop operation the measured ensquared energy is a factor of 2.4 higher than that for the other two AO modes, matching the prediction derived from ensquared energy analysis of the closed-loop and tip-tilt only corrected PSFs shown within Fig.~\ref{fig6}. This figure demonstrates this factor will increase as the spatial sampling is reduced, at the expense of overall transmission e.g. a approximate factor of 5 improvement between closed-loop and the two other AO modes would be achieved if the spatial sampling were reduced to $0.1\arcsec$.

\section{Conclusions}\label{conc}
We have demonstrated the on-sky testing of a three-dimensional integrated optical waveguide device (the photonic dicer) that spatially reformats an AO-corrected telescope PSF into a diffraction-limited pseudo-slit. This component converts any modal noise induced by the atmosphere on the PSF into amplitude and phase variations along the diffraction-limited slit, orthogonal to the dispersion axis in a spectrograph. We have developed all the control and calibration tools required to optimise coupling into a photonic device that is used with an AO system and tested transmission of the photonic dicer across a range of atmospheric and AO operating modes.

Although the improvement in performance that our experimental  system demonstrated is modest, it serves to verify the expected level of performance and is unquestionably useful (a factor of 2.4 in signal, or approximately 1~mag) even in its unoptimised form.

Using the principles demonstrated here, we will be able to improve the performance of spectrographs operating at high spatial resolution near to the diffraction limit using AO.  Even the magnitude increase in depth that our experiment has provided can prove highly valuable. As we have argued, we expect performance to be further improved by refinement of the laser inscription parameters, ensuring that the residual aberrations from the AO system are supported by the modes of the photonic dicer and by optimising the input coupling. We plan to explore these tradeoffs using a combination of theoretical and experimental studies using AO systems both in the laboratory and at the telescope.

It is clear the photonic dicer has the potential to improve the performance of high resolution spectroscopy in areas such as exo-planet detection and characterisation, but we can expect further substantial benefits in these key fields by noting that the current performance of high resolution spectrographs which although already very impressive (100 cms$^{-1}$) is limited ultimately by errors in injecting calibration light using fibres. This is ultimately due to modal noise, where the interference between modes within fibres affect its output intensity pattern, and hence the entrance illumination of the spectrograph. We expect that this problem will be much reduced using a photonic dicer because the reformatted slit is single moded in the dispersion axis, and we expect that variations in the input will not be expressed as variations in the spectrograph line profile.

Aims in future work include; increasing the wavelength range to cover the entire H band; further optimisation of the coupling from CANARY and improved transmission for the photonic dicer through refinements in the manufacturing process.

\section*{Acknowledgments}
R.J.H. and J.R.A-S. gratefully acknowledge support from the Science and Technology Facilities Council (STFC) in the form of a Ph.D studentship (ST/I505656/1) and grant (ST/K000861/1). D.G.M. is supported by an EPSRC Ph.D studentship. R.R.T. gratefully acknowledges funding from the STFC, in the form of an STFC Advanced Fellowship (ST/H005595/1) and through the STFC-PRD scheme (ST/K00235X/1), and also from the Royal Society (RG110551). We wish to thank the European Union for funding via the OPTICON Research Infrastructure for Optical/IR astronomy (EU-FP7 226604). CANARY was supported by Agence Nationale de la Recherche program 06-BLAN-0191, CNRS / INSU, Observatoire de Paris, and Université Paris Diderot Paris 7 in France, STFCl (ST/K003569/1 and ST/I002781/1), University of Durham in UK and European Commission Framework Programme 7 (E-ELT Preparation Infrastructure Grant 211257 and OPTICON Research Infrastructures Grant 226604). Raw data from the experiment is available from the lead authors.




\label{lastpage}


\begin{thebibliography}{}
\bibitem[\protect\citeauthoryear{Betters et al.}{2013}]{Betters2013}Betters C. H., Leon-Saval S. G., Robertson J. G. \& Bland Hawthorn J., 2013. Opt. Express 21, 26103-26112
\bibitem[\protect\citeauthoryear{Bland-Hawthorn et al.}{2010}]{Bland-Hawthorn2010}Bland-Hawthorn J. et al., 2010. Proc. SPIE 7735, 77350N
\bibitem[\protect\citeauthoryear{Bland-Hawthorn et al.}{2011}]{Bland-Hawthorn2011}Bland-Hawthorn J. et al., 2011. Nat. Commun. 2, 581
\bibitem[\protect\citeauthoryear{Birks et al.}{2012}]{Birks2012}Birks T. A., Mangan B. J., Diez A., Cruz J. L. \& Murphy, D. F., 2012. Opt. Express 20, 13996-14008
\bibitem[\protect\citeauthoryear{Cvetojevic et al.}{2009}]{Cvetojevic2009}Cvetojevic N. et al., 2009. Optics Express, vol. 17, issue 21, p. 18643-18650
\bibitem[\protect\citeauthoryear{Cvetojevic et al.}{2012}]{Cvetojevic2012}Cvetojevic N. et al., 2012. Optics Express, vol. 20, issue 3, p. 2062
\bibitem[\protect\citeauthoryear{Chauvin et al.}{2004}]{Chauvin2004}Chauvin G. et al., 2004. A\&A. 425, L29-L32
\bibitem[\protect\citeauthoryear{Cunningham}{2009}]{Cunningham2009}Cunningham C., 2009.  Nat. Photon. 3, 239-241
\bibitem[\protect\citeauthoryear{Davis et al.}{1996}]{Davis1996}Davis K. M., Miura K., Sugimoto N. \& Hirao, K., 1996. Opt. Express 37, 1729-1731
\bibitem[\protect\citeauthoryear{Eisenhauer et al.}{2003}]{Eisenhauer2003}Eisenhauer F. et al., 2003.  ApJ. 597, L121-L124
\bibitem[\protect\citeauthoryear{Gatass \& Mazur}{2008}]{Gatass2008}Gattass R. R. \& Mazur E., 2008. Nat. Photon. 2, 219-225
\bibitem[\protect\citeauthoryear{Gendron et al.}{2011}]{Gendron2011}Gendron E. et al., 2011. A\&A. 529, L2
\bibitem[\protect\citeauthoryear{Ghez et al.}{2003}]{Ghez2003}Ghez A.M. et al., 2003. ApJ. 586, L127-L131
\bibitem[\protect\citeauthoryear{Harris \& Allington-Smith}{2012}]{Harris2012}Harris R. J. \& Allington-Smith J. R.,  2012. MNRAS 428, 3139-3150
\bibitem[\protect\citeauthoryear{Hill}{2008}]{Hill2008}Hill G.J. et al., 2008. Proc. SPIE 7014, 701470
\bibitem[\protect\citeauthoryear{H$\o$g et al}{2000}]{Hog2000}H$\o$g E. et al., 2000. A\&A. 355, L27
\bibitem[\protect\citeauthoryear{Jovanovic et al.}{2012}]{Jovanovic2012}Jovanovic N. et al., 2012. MNRAS. 427, 806-815
\bibitem[\protect\citeauthoryear{Kirkpatrick, Gelatt \& Vecchi}{1983}]{Kirkpatrick1983}Kirkpatrick S., Gelatt C. D. \& Vecchi M. P., 1983. Science 220, 671-680
\bibitem[\protect\citeauthoryear{Kwok}{2004}]{Kwok2004}Kwok S., 2004. Nature 430, 985-991
\bibitem[\protect\citeauthoryear{Lemke et al.}{2011}]{Lemke2011}Lemke U. et al., 2011. MNRAS 417.1, 689-697
\bibitem[\protect\citeauthoryear{Leon-Saval et al.}{2005}]{Leon-Saval2005}Leon-Saval S. G., Birks T. A., Bland-Hawthorn J. \& Englund M., 2005. Opt. Lett. 30, 2545-2547
\bibitem[\protect\citeauthoryear{MacLachlan et al.}{2014}]{MacLachlan2014}MacLachlan D.G. et al., 2014. Proc. SPIE 91511W-91511W-7
\bibitem[\protect\citeauthoryear{Marois et al.}{2008}]{Marois2008}Marois C. et al., 2008.  Science 322, 1348-1352
\bibitem[\protect\citeauthoryear{Myers et al.}{2008}]{Myers2008}Myers R. M. et al., 2008. Proc. SPIE 7015, 70150E
\bibitem[\protect\citeauthoryear{Nasu, Kohtoku \& Hibino}{2010}]{Nasu2005}Nasu Y., Kohtoku M. \& Hibino Y., 2005. Opt. Lett. 30, 723-725
\bibitem[\protect\citeauthoryear{Nolte et al.}{2003}]{Nolte2003}Nolte S., Will M., Burghoff J. \& Tuennermann A., 2003. App. Phys. A. 77, 109-111
\bibitem[\protect\citeauthoryear{Noordegraaf et al.}{2009}]{Noordegraaf2009}Noordegraaf D., Skovgaard P. M., Nielsen M. D. \& Bland-Hawthorn J., 2009. Opt. Express 17, 1988-1994
\bibitem[\protect\citeauthoryear{O'Mahoney}{2003}]{OMahoney2003}O'Mahoney N., 2003. The Newsletter of the Isaac Newton Group of Telescopes 7, 22-24
\bibitem[\protect\citeauthoryear{Spaleniak et al.}{2013}]{Spaleniak2013}Spaleniak I., et al., 2013.   Optics Express, vol. 21, issue 22, p. 27197
\bibitem[\protect\citeauthoryear{Swain et al.}{2010}]{Swain2010}Swain M. R., et al., 2010.  Nature 463, 637-639
\bibitem[\protect\citeauthoryear{Thomson, Kar \& Allington-Smith}{2009}]{Thomson2009}Thomson R. R., Kar A. K. \& Allington-Smith, J. R., 2009. Opt. Express 17, 1963-1969
\bibitem[\protect\citeauthoryear{Thomson et al.}{2011}]{Thomson2011}Thomson R. R., Birks T. A., Leon-Saval S. G., Kar A. K. \& Bland-Hawthorn J., 2011. Opt. Express 19, 5698-5705
\bibitem[\protect\citeauthoryear{Thomson et al.}{2012}]{Thomson2012}Thomson R. R., Harris R. J., Birks T.A., Brown G. Allington-Smith J. R. \& Bland-Hawthorn J., 2012, Opt. Lett. 37, 2331-2333
\bibitem[\protect\citeauthoryear{Vidal et al.}{2014}]{Vidal2014} Vidal F., Gendron, ƒ., Rousset G., Morris, T., Basden A., Myers R., Brangier M., Chemla F., Dipper N., Gratadour, D., Henry D., Hubert Z., Longmore A., Martin O., Talbot, G., and Younger, E. 2014. A\&A. Volume 569, id.A16, 19 pp.
\bibitem[\protect\citeauthoryear{Weitzel et al.}{1996}]{Weitzel1996} Weitzel L., Krabbe A., Kroker H., Thatte N., Tacconi-Garman L. E., Cameron M., and Genzel R., 1996. A\&A. Suppl., 119, 531-546

\end{thebibliography}
\end{document}